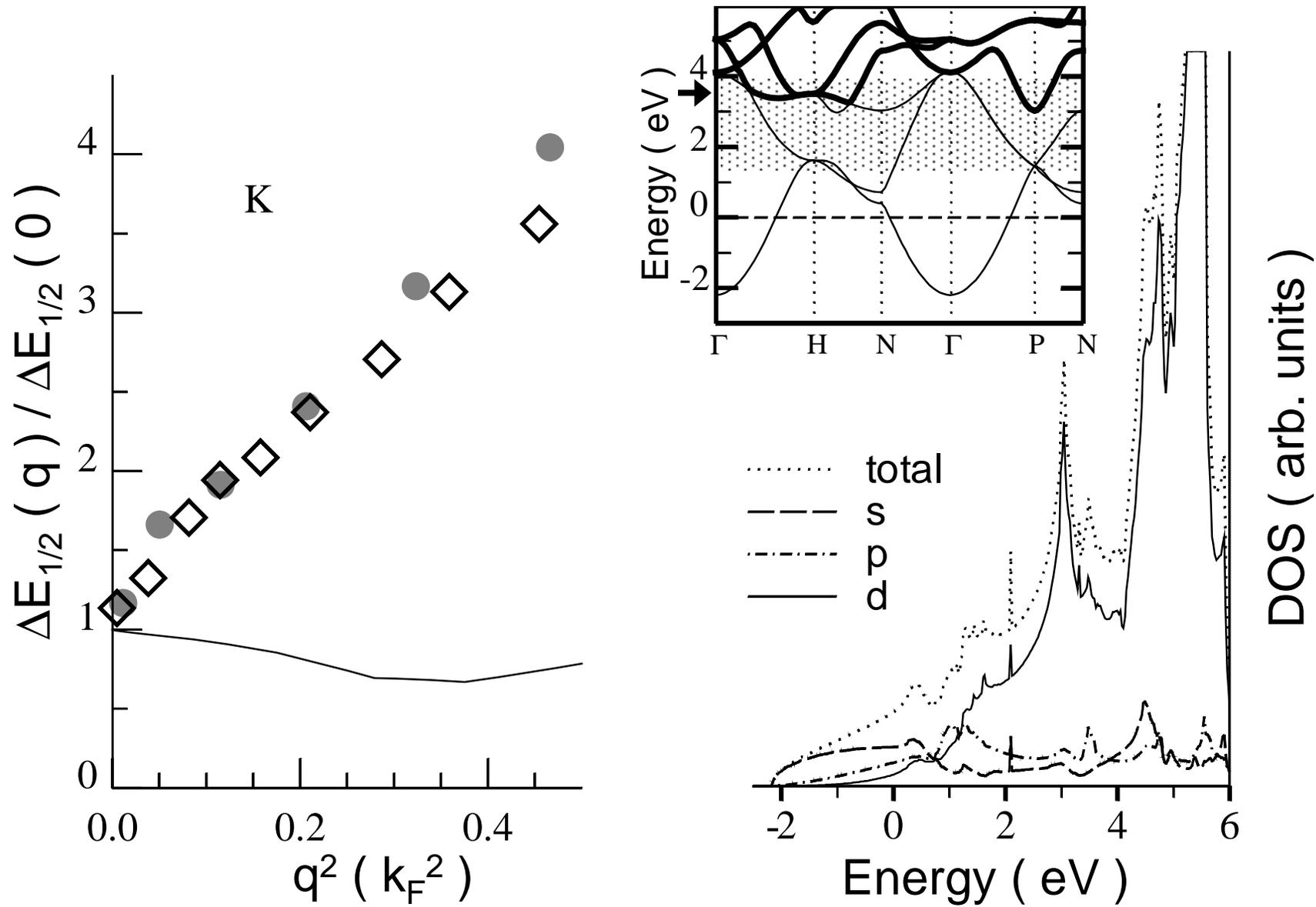

Fig. 1. Eguiluz, Ku & Sullivan

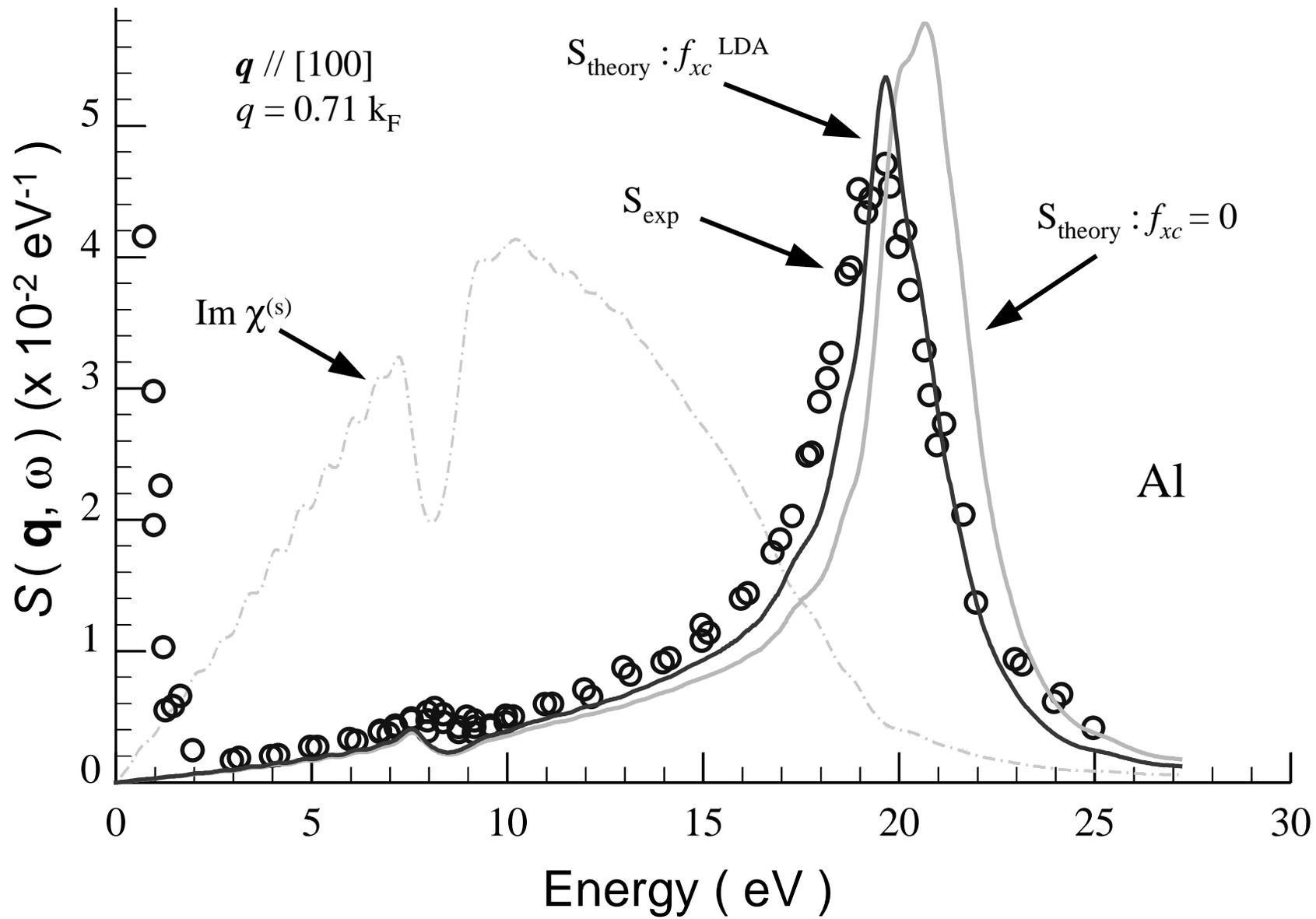

Fig. 2. Eguiluz, Ku & Sullivan

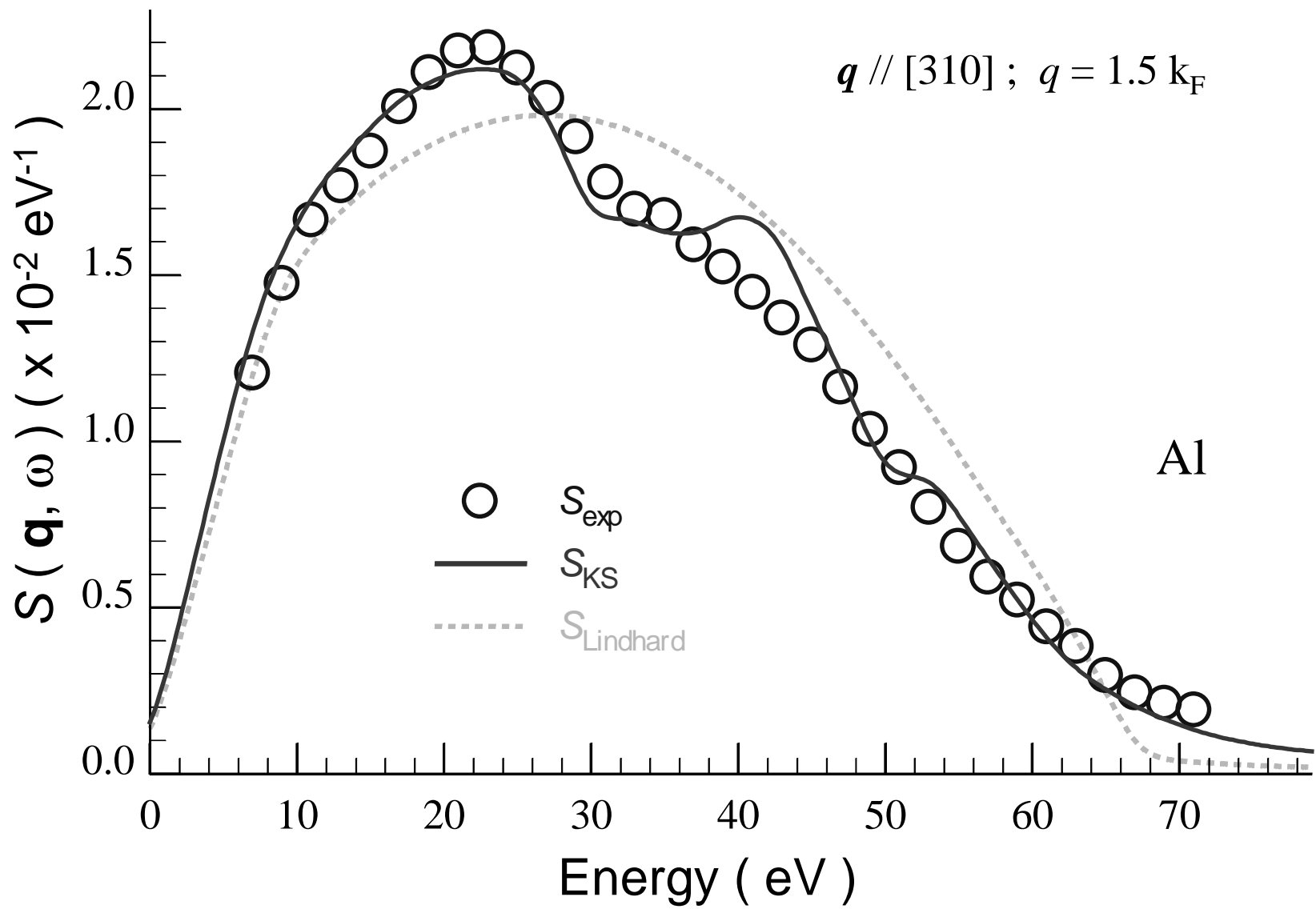

Fig. 3. Eguiluz, Ku & Sullivan

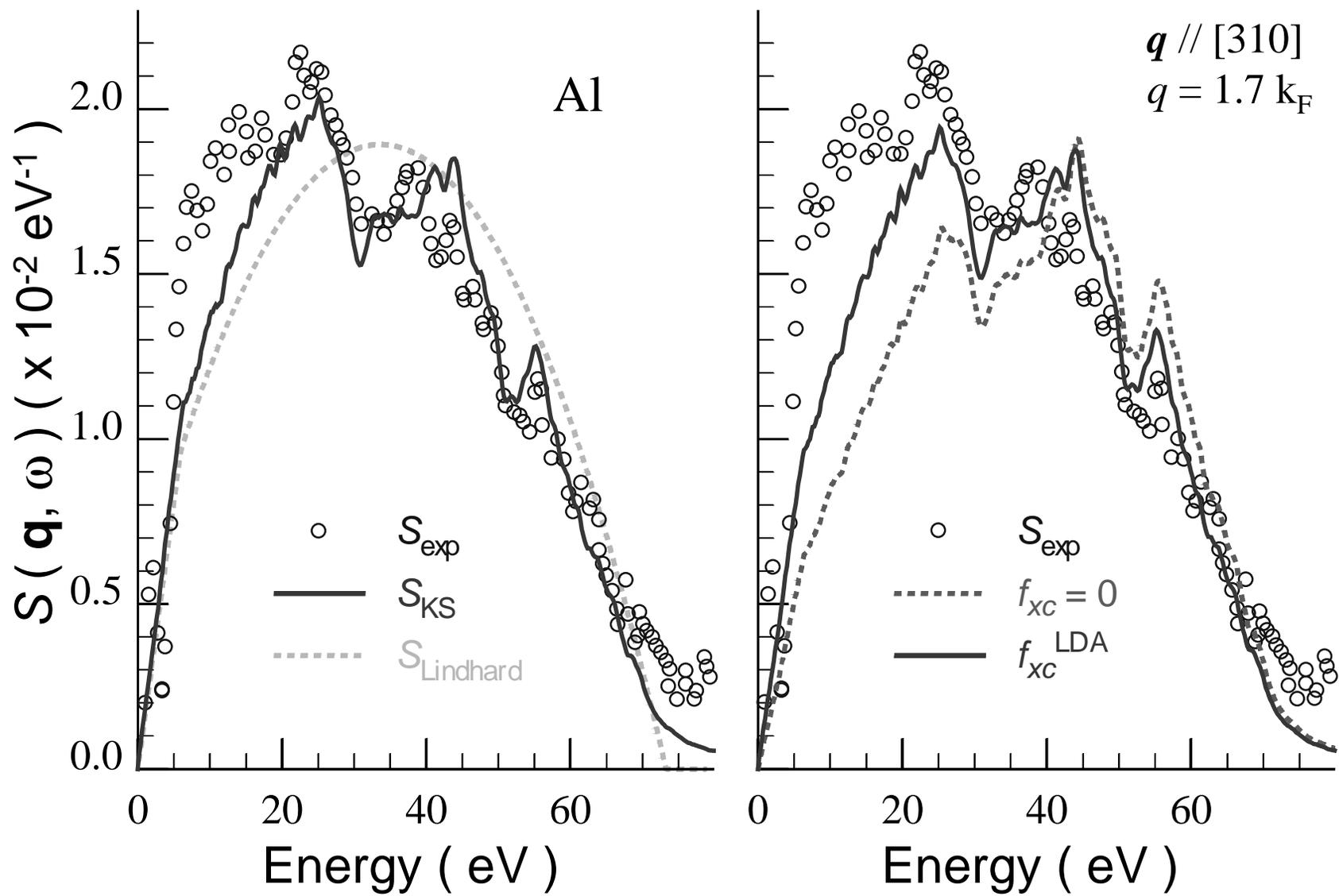

Fig. 4. Eguiluz, Ku & Sullivan

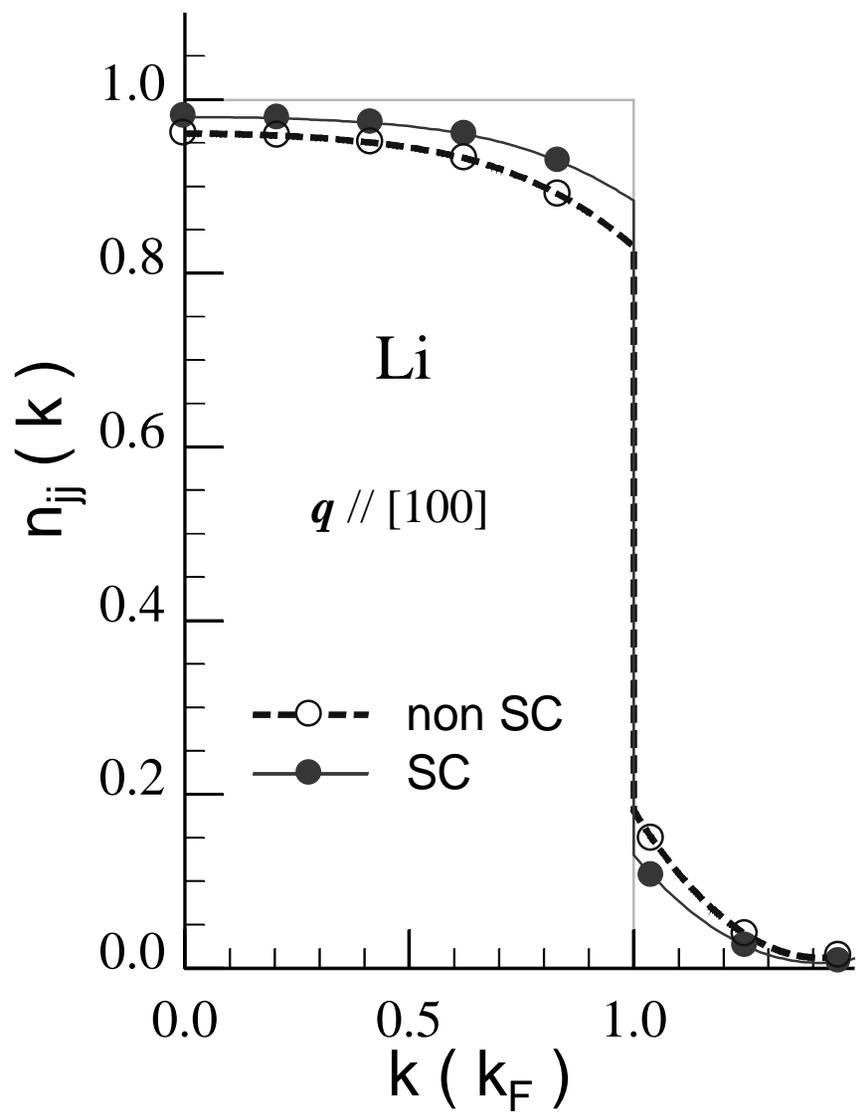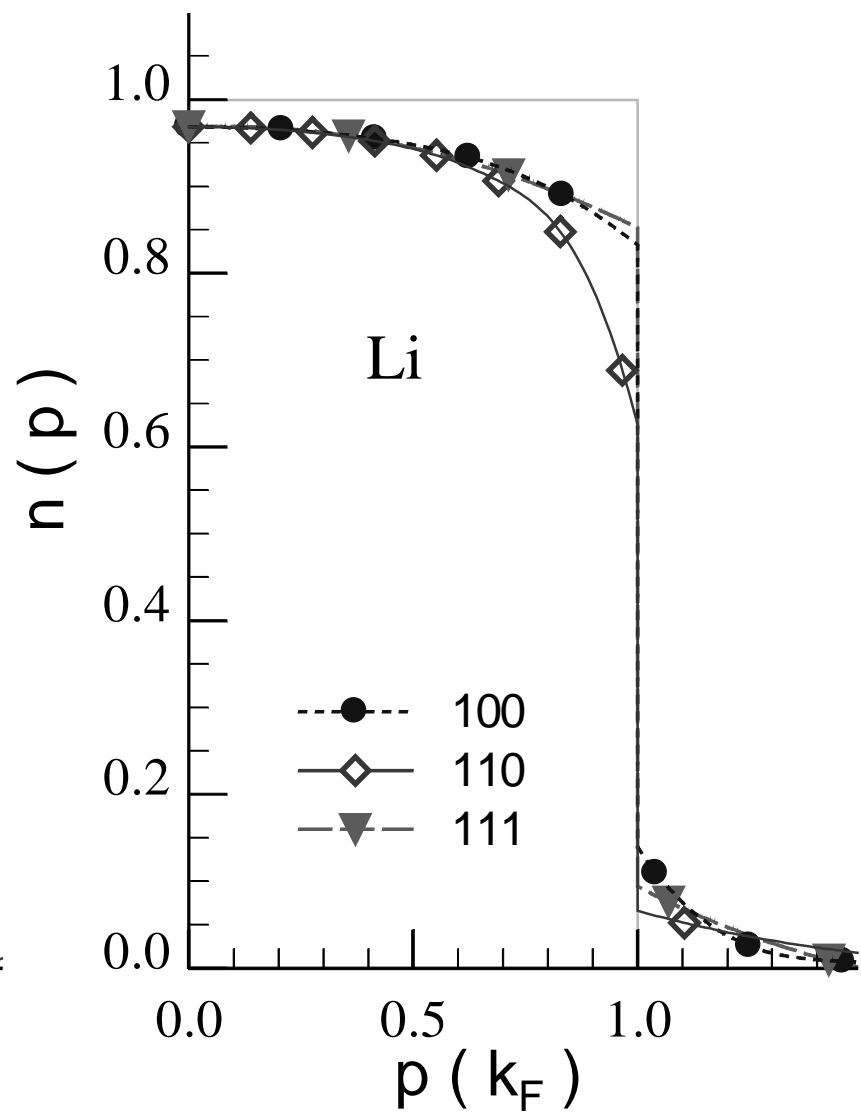

Fig. 5. Eguiluz, Ku & Sullivan

# Dynamical Response of Correlated Electrons in Solids Probed by Inelastic Scattering Experiments:  An *Ab Initio* Theoretical Perspective

Adolfo G. Eguiluz, Wei Ku, and  James M. Sullivan

*Department of Physics and Astronomy, The University of Tennessee, Knoxville, TN 37996-1200*
*and Solid State Division, Oak Ridge National Laboratory, Oak Ridge, TN 37831-6030*

## Abstract

We present results of *ab initio* theoretical investigations of the excitation spectra of correlated electrons in metals (Al, K, and Li) and their interplay with inelastic scattering experiments.  We resolve various anomalies contained in the data, which were originally viewed as signatures of strong dynamical electronic correlations;  we show that, instead, the anomalies are due to band-structure effects.  The underlying theoretical framework in our density-response calculations is time-dependent density-functional theory;  with this scheme we discuss the lifetime of the K plasmon, and the dynamical structure factor of Al.  From a self-consistent solution of the Dyson equation in the *GW* approximation for the electron self-energy, we discuss the electron momentum density in Li. Our results and methods point to a new way of thinking about electronic excitations in real materials. The main challenge ahead is the proper treatment of dynamical correlations for realistic representations of the band structure.

This International Workshop marks the twenty-fifth anniversary of the breakthrough paper by Eisenberger, Platzman and Pandy [1], *"Investigation of X-Ray Plasmon Scattering in Single-Crystal Beryllium,"* which demonstrated the feasibility of probing electronic excitations in condensed matter via (non-resonant) inelastic x-ray scattering spectroscopy (IXSS). Remarkably, the following concepts, quoted from that paper, permeate a large amount of research conducted in the intervening years: *(i)* "We can certainly assert that for electrons in metallic systems the interaction with the lattice produces effects on dispersion and lifetimes which are much larger than many refinements to the random phase approximation (RPA) that have been previously calculated." *(ii)* "At the largest scattering angles studied, the spectrum has a rather peculiar and not understood shape," and *(iii)* "While some of the features in this spectrum no doubt depend on the details of the Be band structure, others may very well be signaling a real breakdown of this simple RPA picture."

While space restrictions do not allow us to review the history of the field as it unfolded in the wake of Ref. [1], it is well documented that *(ii)* and *(iii)* led to an extensive effort to understand the "two-peak" loss structure first observed in Be, and subsequently also in systems with quite different band structures, such as Al, Li, Si, and graphite [2]. On the theoretical side, until quite recently this effort consisted overwhelmingly of work which, statement *(i)* notwithstanding, concentrated on the study of the effects of correlation for electrons in jellium [3-6]. With the advent of synchrotron sources, and the consequent improvement in energy and wave vector resolution (and photon fluxes), plus the availability of better sample preparation techniques, we have witnessed, over the last 10-15 years, major advances on the experimental front [7-10]. However, theoretical understanding of the physics behind the experimental data has lagged behind. Indeed, it seems fair to state that the rather confusing history of various phenomenological "patch ups" of, for the most part, uncontrolled approximations for the many-body problem in jellium, hit a dead end some years ago. It is only recently that *ab initio* calculations —which in essence have the flavor of *(i)*— have begun to be reported [11-14]; with them a new way of thinking about the excitations of correlated electrons in real solids has begun to emerge. Much work remains to be done, though, as we will indicate below.

In this article we offer a theoretical perspective conceived on the basis of selected results of recent *ab initio* studies, and their comparison with experiment. We consider first a case study of plasmon damping. Anomalous behavior of the dispersion of the plasmon *linewidth* in K was observed in electron energy-loss spectroscopy (EELS) experiments [15]. Interestingly, in the absence of *ab initio* theory, the ansatz was made that the explanation of the EELS data may reside in the effects of unspecified dynamical short-range correlations [15]. Note that this proposition is very much in tune with the thrust of the work carried out over the years following the discovery of the "anomalous" two-peak IXSS loss structure in Refs. [1,2]. We show that the plasmon linewidth dispersion in K is, in fact, controlled by plasmon decay into particle-hole pairs involving empty states of *d*-symmetry [16]. We discuss next the dynamical structure factor, $S(\vec{q};\omega)$, of Al using the same theoretical framework. Again, we find that band-structure effects provide a "natural" explanation of the key qualitative feature of the IXSS spectrum for wave vectors for which the plasmon has already Landau-damped [12,13]; specifically, the two-peak structure is induced by a zone-boundary gap which yields an indentation ("excitation gap") in the response function.

Now, by comparison with experiment we infer that, while dynamical many-particle correlations do not play a role in the K "anomaly," they do affect significantly the intensity of the lower frequency "bump" in the two-peak structure in $S(\vec{q};\omega)$ [9,13]. The greatest remaining challenge is a reliable evaluation of these correlations *for band electrons*. We close this article with reference to ongoing work devoted to a first-principles approach to this problem. We present results for the quasiparticle occupation function $n_{jj}(\vec{k})$ and the electron momentum density $n(\vec{p})$ in Li, obtained via a conserving evaluation of the electron self-energy within the *GW* approximation [17]. Our $n_{jj}(\vec{k})$ differs markedly from the one obtained in a recent *GW* calculation [18]; furthermore, our results do not support the "anomalous" quasiparticle weight at the Fermi surface which has been extracted from Compton-scattering measurements [19]. Our results are more in line with new quantum Monte Carlo work [20].

We proceed to sketch the theoretical framework within which our results for $S(\vec{q};\omega)$ are obtained —time-dependent density functional theory (TDDFT) [21]. A general formulation of the time evolution of an interacting electron system in an external potential $v_e(\vec{x},t)$ has been given by Runge and Gross [21]. These authors established the invertibility of the mapping $v_e(\vec{x},t) \to n(\vec{x},t)$, where $n(\vec{x},t)$ is the time-dependent density for the interacting system. From that non-trivial result it is possible to demonstrate that $n(\vec{x},t)$ can be obtained as

$$n(\vec{x},t) = \sum_{v}^{occupied} |\varphi_v(\vec{x},t)|^2 ,  \qquad (1)$$

in terms of the solutions $\varphi_v(\vec{x},t)$ of the time-dependent Kohn-Sham equation,

$$i\hbar \frac{\partial}{\partial t} \varphi_\nu(\vec{x},t) = \left( -\frac{\hbar^2}{2m}\nabla^2 + v_s[n](\vec{x},t) \right) \varphi_\nu(\vec{x},t) ,  \qquad (2)$$

where the single-particle potential $v_s[n](\vec{x},t)$ contains, in addition to the external and Hartree potentials, the exchange-correlation potential $V_{xc}[n](\vec{x},t)$, whose *functional* dependence on the density is implied by our notation. The TDDFT formalism is particularly well suited for the study of the *linear response* of a many-electron system to an external potential $\delta v_e(\vec{x},t)$ [22]. In general, we define the density-response function $\chi$ by the equation $\delta n = \chi \delta v_e$. In TDDFT the linear change in density can also be calculated as $\delta n = \chi^{(s)} \delta v_s$, where $\chi^{(s)}$ is the single-particle response for the *unperturbed* Kohn-Sham system. From Eqs. (1) and (2) $\delta v_s$ can be related to $\delta v_e$, and this leads to the following integral equation for the response function [22]:

$$\chi = \chi^{(s)} + \chi^{(s)} (v + f_{xc}) \chi , \qquad (3)$$

where $v$ is the bare Coulomb interaction, and the exchange-correlation kernel $f_{xc}$ is defined by the equation

$$f_{xc}[n](\vec{x}t; \vec{x}'t') = \frac{\delta V_{xc}[n](\vec{x}t)}{\delta n(\vec{x}'t')} , \qquad (4)$$

where the functional derivative is to be evaluated at the unperturbed density. We emphasize that Eqs. (3) and (4) are formally exact; all explicit effects of dynamical correlations are contained in $f_{xc}$. The spectral representation for the single-particle response $\chi^{(s)}$ in terms of the eigenfunctions and eigenvalues of the unperturbed Kohn-Sham problem is of the usual form. For a periodic crystal it is convenient to work with the Fourier transform of $\chi^{(s)}$, given by the equation

$$\chi^{(s)}_{\vec{G},\vec{G}'}(\vec{k};\omega) = \frac{1}{V} \sum_{\vec{k}'}^{BZ} \sum_{j,j'} \frac{f_{\vec{k}',j} - f_{\vec{k}'+\vec{k},j'}}{E_{\vec{k}',j} - E_{\vec{k}'+\vec{k},j'} + \hbar(\omega+i\eta)} \langle \vec{k}',j | e^{-i(\vec{k}+\vec{G})\cdot\hat{\vec{x}}} | \vec{k}'+\vec{k},j' \rangle$$

$$\times \langle \vec{k}'+\vec{k},j' | e^{i(\vec{k}+\vec{G}')\cdot\hat{\vec{x}}} | \vec{k}',j \rangle , \qquad (5)$$

where $V$ is the normalization volume, $\vec{G}$ is a vector of the reciprocal lattice, $j$ is a band index, and all wave vectors are in the first Brillouin zone (BZ). We evaluate Eq. (5) from Kohn-Sham states and eigenvalues obtained using the local-density approximation (LDA) in the evaluation of the exchange-correlation potential $V_{xc}(\vec{x})$. In the above Fourier representation, Eq. (3) is turned into a matrix equation that we solve numerically. The dynamical structure factor is given by (we omit well-known factors) $S(\vec{q};\omega) \approx \mathrm{Im}\,\chi_{\vec{G},\vec{G}}(\vec{k};\omega)$, where $\vec{q} = \vec{k}+\vec{G}$; in this equation the *wave-vector transfer* $\vec{q}$ is not restricted to the BZ.

The TDDFT linear-response framework proves useful for the purpose of sorting out the effects of the one-particle band structure from those of dynamical many-body correlations, as the latter can be turned off by setting $f_{xc} = 0$. Although this "sorting out" is not always possible (i.e., these effects may be intrinsically intertwined in systems with more complex electronic structures than the ones considered in this article), we provide next two important examples where this viewpoint turns out to be rewarding.

Figure 1 (left panel) shows a well-converged calculation [16,23] of the plasmon linewidth dispersion of K (solid circles); we have set $f_{xc} = 0$, aiming at elucidating the impact of single-particle decay channels. Since the $3p^6$-derived core states play a role in the plasmon dynamics of K, we evaluate $\chi^{(s)}$ using the full-potential linearized augmented plane wave (LAPW) method [24]. The calculated full-width at half-maximum of the plasmon peak, $\Delta E_{1/2}(\vec{q})$, is compared with the EELS data of vom Felde et al. [15] (empty diamonds). Clearly, our results are in excellent agreement with experiment; since this agreement is obtained for $f_{xc} = 0$, we conclude that the plasmon linewidth *dispersion* of K is *not* controlled by a dynamical many-body mechanism.

Our result is striking, as intuitive expectations based on the fact that the gap just above the Fermi surface at the *N*-point is small yield the result shown by the solid line (left panel), which corresponds to an evaluation of the dielectric function to second order in an empirical pseudopotential [25]. It is apparent that the use of nearly-free-electron states and eigenvalues in the evaluation of $\chi^{(s)}$ breaks down. We explain this breakdown by analyzing the impact of key "final state" bands, shown in the inset of Fig. 1, in which the shaded strip is the $\omega$-interval representing all the single-particle states which may couple to the plasmon (as determined by the conservation laws of energy and crystal momentum) [26]. Keeping just the first three valence bands (thin solid lines in the inset), the calculated $\Delta E_{1/2}(\vec{q})$ turns out to agree well with the solid line in the left panel of Fig. 1 [16]; this is reasonable, as the states kept are, for the most part, nearly-free-electron-like. When three additional bands (thick solid lines in the inset) are included in the evaluation of $\chi^{(s)}$, the plasmon linewidth dispersion curve changes *qualitatively*, and the calculated $\Delta E_{1/2}(\vec{q})$ agrees well with the EELS data (on the relative scale of Fig. 1) [16]. Thus, *these three bands provide the key decay channels for the plasmon of K*. It is crucial that the bands in question are of *d*-character —cf. the angular momentum-resolved density of states (DOS) shown on the right panel of Fig. 1; the effect of these bands cannot be approximated by nearly free-electron states.

We emphasize that, although the single-particle states entering Eq. (5) do not have the meaning of quasiparticle states, the only approximation made in the evaluation of $\chi^{(s)}$ is the LDA. We have performed additional calculations [16] within the random-phase approximation (RPA), in which *all* exchange-correlation effects are left out, including those in the band structure, which now corresponds to the Hartree approximation. The key physical change is that the flat bands are shifted upwards in relation to the LDA *d*-bands of Fig. 1; the Hartree *d*-bands lie almost entirely *above* the shaded strip, and, as a result, $\Delta E_{1/2}(\vec{q})$ now differs significantly from experiment —it has a much smaller slope than the data [16]. Thus, the correlations contained in the one-particle-like Kohn-Sham states of *d*-symmetry play a non-trivial role, via $\chi^{(s)}$, in the explanation of the plasmon damping mechanism. These correlations are usually not dealt with explicitly in many-body models of interacting electrons.

We turn next to a discussion of the physics of the dynamical structure factor of Al, and its interplay with the IXSS data. We base our analysis on the same TDDFT linear-response framework introduced above. Let us stress that the explanation of the "anomalous" dispersion of the plasmon linewidth in K was made possible by a "fortunate" circumstance: the Kohn-Sham single-particle response $\chi^{(s)}$ contained enough of the physics of the problem that we were close to the experimental data (or "the exact answer") upon setting $f_{xc} = 0$ in Eq. (3). A similar situation is only partially realized in the present case —yet, from $\chi^{(s)}$ we are able to identify the main qualitative feature of the spectrum for large wave vectors.

In Fig. 2 the experimental [9] $S(\vec{q};\omega)$ for Al (labeled $S_{\exp}$) for $|\vec{q}| = 0.71 k_F$ ($\vec{q}$ along the (100) direction) is compared with spectra calculated from Eq. (3) [27]. It is important to note at the outset that the theoretical and experimental spectra are plotted in *absolute units*, by use of the *f*-sum rule; there is no lining up of peak hights in any of our results. First we note that $\text{Im}\,\chi^{(s)}$ bears no immediate relationship with the main loss peak lying at about 20 eV. This is understandable; for the present, rather small (in the IXSS sense) wave vector the dominant spectral feature is the plasmon, which is about to enter the Landau-damping regime, as evidenced by the pronounced low-frequency tail in the spectrum; this collective mode is outside the realm of the single-particle response $\chi^{(s)}$. In this wave vector domain the RPA-like response function $\chi$ obtained from Eq. (3) for $f_{xc} = 0$ shares the usual RPA "luck," and contains enough of a renormalization of the single-particle response $\chi^{(s)}$ due to the presence of the bare Coulomb interaction $v$ in Eq. (3) that it leads to an $S(\vec{q};\omega)$ (labeled $f_{xc} = 0$ in Fig. 2) which accounts for the main loss reasonably well. But we can do better, if we invoke an adiabatic ansatz [28], and introduce in Eq. (3) the short-range correlation effect of $f_{xc}$ within the LDA. As seen in Fig. 2, this approximation for $S(\vec{q};\omega)$ (labeled $f_{xc}^{\text{LDA}}$) improves on the $f_{xc} = 0$ result with regard to both the position of the plasmon peak, which is shifted downward (an effect of the exchange-correlation hole surrounding each electron), and the overall lineshape.

It is important to notice the feature present in the measured spectrum in Fig. 2 at ~ 8 eV. The same clearly correlates with the indentation observed in $\text{Im}\,\chi^{(s)}$ at about the same energy in this otherwise Lindhard-like single-particle response. This "quasi-gap" in $\text{Im}\,\chi^{(s)}$ is traced to the periodic-potential-induced gap in the band structure of Al at the (100) zone boundary.

In Fig. 3 we consider the loss spectrum for a larger wave vector transfer, $|\vec{q}| = 1.5 k_F$ ($\vec{q}$ is along the (310) direction). The collective mode has now Landau-damped, and the response is incoherent. The overall shape of the measured loss spectrum $S_{exp}$ [9] is usually described as one-peak, one-shoulder, structure; the origin of this shape is given below, in the context of Fig. 4, in which the "shoulder" is more pronounced. Remarkably, the $S(\vec{q};\omega)$ obtained directly from $\text{Im}\,\chi^{(s)}$ —labeled $S_{KS}$— reproduces $S_{exp}$ very well. In view of the exact nature of Eq. (3), this can only be the case if in the $(\vec{q};\omega)$ region relevant to Fig. 3 the many-body kernel $f_{xc}$ basically "cancels" the effect of the bare Coulomb interaction. This figure thus contains a strong hint for what a fully *ab initio* theory of the many-body correlations contained in $f_{xc}$ should be able to predict. (The bump in the calculated spectrum at $\omega \sim 40\,\text{eV}$ would presumably not be as pronounced in the presence of dynamical effects left out in the calculation.)

Finally, in Fig. 4 we display data for $S(\vec{q};\omega)$ (already published in Ref. [13]) for the regime in which the incoherent loss spectrum is dominated by the "anomalous" two-peak structure. (In the present case $S_{exp}$ is taken from Ref. [8]; the momentum transfer is again along the (310) direction.) Consider first the left panel. It is apparent that, as was the case in Fig. 3, the single-particle response $\text{Im}\,\chi^{(s)}$ yields an $S_{KS}$ which reproduces the main features of $S_{exp}$ quite well. In order to understand this result, it is helpful to compare $S_{KS}$ with its counterpart for jellium, the Lindhard function (labeled $S_{Lindhard}$), also shown in the figure; the key difference between both loss functions stems from the indentation in $\text{Im}\,\chi^{(s)}$ at $\sim 32$ eV, which correlates well with a similar feature observed in $S_{exp}$. Since the (310) direction differs from the (100) direction by a relatively small angle, this "indentation" is partially traced to the large gap in the Al band structure at the (200) zone boundary.  Further details are given by Larson in these Proceedings; see also Ref. [12].  Thus, we are in the presence of a "quasi-gap" in the excitation spectrum contained in $\mathrm{Im}\,\chi^{(s)}$.  It is important to realize that this feature has the same origin as the one observed in Fig. 2 for $\omega \approx 8\,\mathrm{eV}$.

Now, with reference to the right panel of Fig. 4, once we turn on the Coulomb interaction $v$ in Eq. (3) (curve labeled $f_{xc} = 0$, this is an "RPA-like" response), the calculated spectrum worsens considerably, relative to experiment —the intensity on the low-energy side of the spectrum is far too low.  This is an indication of the importance of the many-body kernel $f_{xc}$, which, as we did before (Fig. 2), we evaluate self-consistently with the electronic structure within the LDA (curve labeled $f_{xc}^{\mathrm{LDA}}$).  This inclusion of short range correlations leads to a marked improvement in the intensity of the low-energy "peak."  Note that, of course, the "excitation gap" built into $\mathrm{Im}\,\chi^{(s)}$ is reflected in both theoretical spectra shown on the right panel.

We summarize the above discussion by emphasizing that the predominant two-peak loss structure (Fig. 4), or one-peak, one-shoulder loss structure (Fig. 3), can pictorially be described as an indentation "carved out" from the Lindhard function because of an excited-state gap in the Kohn-Sham band structure.  We reiterate the sound basis which TDDFT gives to this argument, since the Kohn-Sham response $\chi^{(s)}$ enters Eq. (3) "legally," even if the individual terms in the summation performed in Eq. (3) do not have the meaning of quasiparticle states.

From the theoretical point of view, we still have a real challenge:  the evaluation of a dynamical many-body kernel $f_{xc}$.  The above comparison with experiment has provided interesting benchmarks as to what one would really like to *predict*. It should be apparent by now that we are referring to an evaluation of $f_{xc}$ for band electrons, since starting from jellium, as has traditionally been done, is far off the final answer. A great virtue of the TDDFT scheme is that it gives us a starting point, $\chi^{(s)}$, which already contains a good deal of the physics of the real system (recall quote *(i)* at the outset of this paper!). Now, to the best of our knowledge, at the present time there are no systematic techniques within DFT to produce time-dependent exchange-correlation functionals from which successively better approximations for frequency-dependent $f_{xc}$'s could be evaluated and judged by comparison with, say, IXSS data. Thus, we pursue a diagrammatic alternative.

A rigorous formulation of the many-body problem of interacting electrons starts out from the Dyson equation for the one-particle Green's function, which we write down as

$$G^{-1}(1,1') = G_{LDA}^{-1}(1,1') - \tilde{\Sigma}(1,1') , \qquad (6)$$

where $G_{LDA}(1,1')$ is the Green's function for "free" propagation in the LDA band structure, and the labels $1,1'$ denote space-time points; the time variables are Matsubara times $0 \leq \tau, \tau' \leq \beta\hbar$. All correlations beyond the LDA are contained in the self-energy $\tilde{\Sigma}(1,1')$, which is a functional of $G$ —thus the self-consistent nature of the problem. In order to avoid double-counting the interactions already built into $G_{LDA}(1,1')$, we define the self-energy according to the equation $\tilde{\Sigma}(1,1') = \Sigma(1,1') - (V_H(1) + V_{xc}(1))\delta(1-1')$, where $V_H(1)$ and $V_{xc}(1)$ are, respectively, the Hartree and exchange-correlation (XC) potentials entering the Kohn-Sham equation in LDA; $\Sigma(1,1')$ is the self-energy functional usually defined in textbook "empty lattice" formalisms.

We close this article by presenting preliminary self-consistent results for the quasiparticle occupation function and the momentum density for interacting electrons in Li metal; the latter quantity is indirectly probed in IXSS experiments performed in the Compton limit [29]. The motivation for this calculation is that recent Compton measurements by Schülke *et al.* [19] have suggested that the value of the quasiparticle weight at the Fermi surface, $Z_{k_F}$, is anomalously low. This suggestion would imply that the electronic states near the Fermi surface of Li are strongly correlated, and/or that large and novel effects of the crystal structure may be at play. Moreover, recent *GW*-based calculations by Kubo [18] suggest that, although $Z_{k_F}$ is somewhat larger than the value extracted from the experiments of Ref. [19], it is nonetheless substantially smaller than "standard" predictions for electrons in jellium with the bulk density of Li. In addition, a rather substantial directional dependence of the quasiparticle occupation function was found in Ref. [18].

We work within the *GW* approximation due to Hedin [30], in which the exchange-correlation contribution to the self-energy functional is given as

$$\Sigma_{xc}(1,1') = -V_S(1,1')\,G(1,1')\,, \tag{7}$$

where the shielded (or screened) interaction is the solution of the integral equation

$$V_S(1,1') = v(1-1') + \int d\overline{1}\int d\overline{2}\; v(1-\overline{1})\, P(\overline{1},\overline{2})\, V_S(\overline{2},1')\,, \tag{8}$$

where $P(1,1') = 2\,G(1,1')\,G(1',1)$ is the polarizability. (Note: the full self-energy contains, in addition to Eq. (7), the standard Hartree contribution; also, in the present case the polarizability *P* neglects vertex corrections, which is the spirit of Hedin's approximation.) We stress that Eqs. (6)-(8) must be solved self-consistently. As shown by Baym and Kadanoff [31], this self-consistency, coupled with the structure of the chosen self-energy functional, guarantees that microscopic conservation laws are fulfilled *exactly* ("conserving" approximation).

A convenient method of solution of the above set of equations is put forth in Ref. [17]. We work in the Bloch basis of the Kohn-Sham states $\varphi_{\vec{k}j}(\vec{x})$, and obtain the Green's function $G_{jj}(\vec{q};i\omega_n)$. From it, we obtain the occupation function $n_{jj}(\vec{k})$ for quasiparticle states, labeled by the quantum numbers $\vec{k}, j$ of the Bloch states,

$$n_{jj}(\vec{k}) = \frac{1}{\beta\hbar} \sum_{i\omega_n} e^{i\omega_n 0^+} G_{jj}(\vec{k}, i\omega_n) , \qquad (9)$$

in terms of which we can construct the electron momentum density $n(\vec{p})$ according to the equation

$$n(\vec{p}) = \sum_j \phi^*_{\vec{k}j}(\vec{p}) n_{jj}(\vec{k}) \phi_{\vec{k}j}(\vec{p}) , \qquad (10)$$

where the $\{\phi_{\vec{k}j}(\vec{p})\}$ are the Fourier coefficients of the Bloch states, $\vec{p} \equiv \vec{k} + \vec{G}$, and we have assumed that the occupation function is diagonal in the band indices. We note that for a periodic crystal, since momentum is not a good quantum number, the momentum density given by Eq. (10) is not the same as the quasiparticle occupation function defined by Eq. (9); both functions are identical for interacting electrons in jellium.

As can be seen in the left panel of Fig. 5, and also in Table I, self-consistency has the effect of *increasing* the value of $Z_{k_F}$. This effect (which may depend on the structure of the *GW* self-energy) was already observed in calculations for jellium [17,32], K, and Si [17], and originates in the reduction of the weight of the many-body satellites in the spectral function as the propagators are dressed with the self-energy. It is apparent that our result for $n_{jj}(\vec{k})$ ($j$=1 corresponds to the only occupied band in Li) is not indicative of anomalously large correlations at the Fermi surface. This conclusion is in qualitative disagreement with the (non self-consistent) *GW* results of Kubo, in which the jump discontinuity along the three high symmetry directions is a factor of 2-4 smaller than our corresponding results. Of course, our calculated value of $Z_{k_F}$ disagrees even more with the vanishingly small jump extracted from the Compton measurements of Ref. [19]. Whether short range correlations (not included in the *GW* diagram) may account for this discrepancy is unknown; however, it appears unlikely for this to be the case. This situation will be discussed in more detail elsewhere.

In the right panel of Fig. 5 we present results for the electron momentum density $n(\vec{p})$, which is the basic building block for the evaluation of the Compton profile. This quantity does show some directional dependence —the jump at the Fermi surface, $\Delta n(p)$, is smallest for the (110) direction. This modest amount of anisotropy is a *lattice effect*, which we traced to the impact of the (110) Fourier coefficient of the crystal potential (which is responsible for the large gap at the $N$-point) on the coefficients $\phi_{\tilde{k}j}(\vec{p})$ which enter Eq. (10). A related effect is that secondary Fermi surfaces become more important for the (110) direction, and show up as features in the momentum density for wave vectors beyond $k_F$.

The preceding discussion can be summarized by noting that our results for the quasiparticle occupation function, and the related electron momentum distribution, do not support the notion that there is some exciting anomaly in the physics of the electrons at the Fermi surface of Li. Our results are more in line with those obtained by Takada and Yasuhara [33] for Li-jellium (see Table II), and with the very recent quantum Monte Carlo calculations of Filippi and Ceperley [20].

We thank Ben Larson and Andrzej Fleszar for many helpful discussions on the subject of the dynamical structure factor of Al, and Wolf-Dieter Schöne for help with the quasiparticle calculations. This work was supported by NSF Grant No. DMR-9634502 and the National Energy Research Supercomputer Center. ORNL is managed by Lockheed Martin Energy Research Corp. for the Division of Materials Sciences, U.S. DOE under contract DE-AC05-96OR2464.

TABLE I. Lithium: Calculated value of the quasiparticle renormalization factor at the Fermi surface, $Z_{k_F}$, and calculated value of the jump in the electron momentum density $\Delta n(p)$ at the Fermi surface.

| Direction | $Z_{k_F}$ | | | $\Delta n(p)$ | | |
|---|---|---|---|---|---|---|
| | 1st Evaluation | 1st Iteration | Converged | 1st Evaluation | 1st Iteration | Converged |
| [100] | 0.63 | 0.70 | 0.72 | 0.61 | 0.69 | 0.71 |
| [110] | 0.62 | 0.69 | 0.72 | 0.50 | 0.56 | 0.57 |
| [111] | 0.63 | 0.70 | 0.72 | 0.70 | 0.76 | 0.77 |

| Direction | $Z_{k_F}$ | | | |
|---|---|---|---|---|
| | Present Work | Kubo Ref. [18] | Schülke et al. Ref. [19] | Takada and Yasuhara, Ref. [33] |
| [100] | 0.72 | 0.35 | 0.1 ± 0.1 | |
| [110] | 0.72 | 0.15 | | 0.67 |
| [111] | 0.72 | 0.25 | | |

Table II. Li: Comparison of values of $Z_{k_F}$ obtained in the present work (cf. Table I) with values published in the literature.

FIGURE CAPTIONS

Fig. 1.  Left panel: Plasmon linewidth dispersion for K. Comparison of our theoretical results (solid circles) with the EELS data of Ref. [15] (diamonds), and the theoretical results of Ref. [25] (solid line). Theory is for (110) propagation; the EELS data are for polycrystalline K. Right panel: Calculated DOS for K —total DOS and contributions from states of $s,p$, and $d$- symmetry [24]; the zero of energy is the Fermi level. Inset: LDA band structure of K; the arrow indicates the value of $\omega_p(0)$ (see text and [26]).

Fig. 2.  Comparison of measured IXSS spectrum of Al [9] (empty circles) for a wave vector transfer $|\vec{q}| = 0.71 k_F$ ($\vec{q}$ is along the (100) direction) with calculations based on the solution of Eq. (3) for $f_{xc} = 0$ (no dynamical correlation effects), and $f_{xc}$ evaluated in the LDA. Also shown is Im $\chi^{(s)}$. See text.

Fig. 3.  IXSS spectrum of Al [9] (empty circles) for a wave vector transfer $|\vec{q}| = 1.5 k_F$ ($\vec{q}$ is along the (310) direction), compared with the loss function obtained from Kohn-Sham electrons (solid line), and with its counterpart for electrons in jellium (the Lindhard function). See text.

Fig. 4.  Left panel: IXSS spectrum of Al [8] (empty circles) for a wave vector transfer $|\vec{q}| = 1.7 k_F$ ($\vec{q}$ is along the (310) direction), compared with the loss function obtained from Kohn-Sham electrons (solid line), and with its counterpart for electrons in jellium (the Lindhard function). Right panel: The same IXSS data are compared with the $S(\vec{q};\omega)$ obtained from the solution of Eq. (3) for $f_{xc} = 0$ (no dynamical correlation effects), and $f_{xc}$ evaluated in the LDA. See text.

Fig. 5. Left panel: Quasiparticle occupation function $n_{jj}(\vec{k})$ for correlated electrons in Li, calculated in the *GW* approximation. Shown are the results for the initial evaluation of the one-particle Green's function, and for the fully self-consistent solution of the Dyson equation (6). Right panel: Electron momentum density $n(\vec{p})$ for the three high-symmetry directions in Li. To facilitate comparison with previous work [18,19], the results shown are for the first solution of Eq. (6).